\documentclass[11pt,twoside,a4paper]{article}

\usepackage[american]{babel}
\usepackage{bm}
\usepackage{amsfonts}
\usepackage{graphicx}
\usepackage{amsmath}
\usepackage{amssymb}
\usepackage{amsthm}
\usepackage{a4wide}

\newtheorem{theorem}{Theorem}
\newtheorem*{remark}{Remark}

\newcommand{\BE}{\begin{equation}}
\newcommand{\EE}{\end{equation}}

\numberwithin{equation}{section}

\newcommand{\cE}{\mathcal{E}}

\newcommand{\cO}{\mathcal{O}}

\newcommand{\mR}{\mathbb{R}}
\newcommand{\e}{\mathrm{e}}

\newcommand{\deltaV}{{\delta V}}
\newcommand{\vF}{v_F}

\newcommand{\jj}{\boldsymbol{j}}

\newcommand{\xx}{{\boldsymbol{x}}}
\newcommand{\pp}{{\boldsymbol{p}}}

\newcommand{\vv}{\boldsymbol{v}}

\newcommand{\ssg}{\boldsymbol{\sigma}}

\newcommand{\bk}[1]{{\langle #1 \rangle}}
\newcommand{\abs}[1]{{\vert {#1} \vert}}

\newcommand{\pt}{\partial}

\newcommand{\feq}{F}

\newcommand{\refl}{{\sim}}

\DeclareMathOperator{\DIV}{div}

\usepackage{color}
\usepackage{comment}

\begin{document}

\begin{center}
\Large{{\bf Mathematical modelling of charge transport \\ in graphene heterojunctions}}
	\\[.5cm]
	\small{L. Barletti,\footnote{Dipartimento di Matematica e Informatica ``U. Dini'', Universit\`a di Firenze} 
	\quad G. Nastasi,\footnote{Dipartimento di Matematica e Informatica, Universit\`a di Catania} 
	\quad C. Negulescu,\footnote{Institut de Math\'ematiques de Toulouse, Universit\'e Paul Sabatier, Toulouse} 
	 \quad V. Romano\footnote{Dipartimento di Matematica e Informatica, Universit\`a di Catania} }	\\
 \end{center}
\begin{abstract}
A typical graphene heterojunction device can be divided into two classical zones, where the transport is basically diffusive, 
separated by a ``quantum active region'' (e.g., a locally gated region), where the charge carriers are scattered according to the laws of 
quantum mechanics.
In this paper we derive a mathematical model of such a device, where the classical regions are described by drift-diffusion equations and
the quantum zone is seen as an interface where suitable transmission conditions are imposed that take into account the 
quantum scattering process.
Numerical simulations show good agreement with experimental data.
\end{abstract}

\section{Introduction}
\label{S1}
Graphene-based electronics has been the subject of an intensive theoretical and experimental research since the discovery of 
this striking two-dimensional material in 2004 \cite{NG2004}.
%
Particularly promising are device architectures that mimic the ordinary semiconductor heterojunctions between positively and negatively doped regions (``n''  and ``p'' regions, 
respectively). 
In graphene, such junctions can be obtained by suitable gates configurations, since the electron/hole density can be locally tuned by local electrostatic fields 
\cite{CF2006,Fang07,Huard2007,YK2009}.  
Thus, a sharp potential variation corresponds to a transition between two differently doped regions, so that we speak of ``p-n'' or ``n-p'' heterojunctions. 
In correspondence of such junctions very interesting phenomena occur, such as the so-called Klein paradox \cite{KNG2006,YK2009} and the negative refraction 
(Veselago) electron lensing \cite{CFA2007,LeeParkLee2015}, which could be exploited to create innovative devices.
\par
The aim of this paper is to apply the theory of diffusive quantum transmission conditions, developed in Ref.\ \cite{BN2018}, 
to the mathematical modelling of a device of this kind. 
\par
The concept of quantum interface conditions goes back to the work of Ben Abdallah and coworkers \cite{NBA98,NBA02}, 
and was initially developed in the framework of kinetic equations. 
The  corresponding diffusion theory has been obtained in Refs.\ \cite{DS98,DEA02}, where a boundary layer analysis leads to diffusive transmission conditions.
Such conditions permit to link two ``classical regions'', described by classical (or semiclassical) drift-diffusion equations,  separated by a 
localized quantum interface (e.g.\ a sharp potential variation), which scatters electrons according to the laws of 
quantum mechanics.
In particular, the transmission conditions contain a parameter, dubbed ``interpolation coefficient'' (by analogy with the ``extrapolation coefficient'' occurring in neutron transport theory \cite{BBS1984}), that depends on the scattering coefficients, thus containing the quantum information of the dynamics at the interface.
\par
The theory has been recently revisited in the case of graphene in Ref.\ \cite{BN2018}. 
The main novelty in such a case comes from the fact that electrons in graphene feature a conical intersection between the conduction band 
and the valence band and, therefore, the behaviour of charge carriers is well described by a Dirac-like equation \cite{CastroNeto09}.
This fact, with respect to traditional semiconductors, not only changes the dispersion relation from quadratic to linear but also introduces a stronger coupling between 
positive-energy and negative-energy electrons (the latter to be described as holes)
As we shall see, the populations of electrons and holes are independent in the classical regions but become (in general) coupled by the
quantum interface, so that the interpolation coefficient becomes an interpolation matrix.
\par
Clearly, the theory of quantum interfaces is very attractive when dealing with graphene heterojunctions since, 
as we have remarked at the beginning of this introduction, they originate the most interesting quantum effects.
For heterojunction devices, therefore, quantum transmission conditions may represent a useful tool for modelling purposes.
The present paper is exactly aimed at illustrating the potentiality of this approach in the case of a prototypical n-p-n graphene device \cite{Huard2007,Osyilmaz2007,YK2009}.
\par
Let us present now the outline of this paper.
In Section \ref{S2} we review the main results of Ref.\  \cite{BN2018}. 
In particular, we show how the densities at both sides of a quantum interface are connected by diffusive transmission conditions (Theorem \ref{T2}). 
Such conditions depend on ``asymptotic densities'' associated to the solution of a four-fold Milne (half-space) kinetic problem which, in turn, arises from a boundary layer
analysis involving quantum reflection and transmission coefficients.
A result about existence and other properties of such asymptotic densities is summarized in Theorem \ref{T1}. 
\par
The solution of the Milne problem represents a surviving kinetic step, that one would like to avoid when working in a diffusion framework.
Then, it is natural to look for some approximations that allow to write down explicitly the solution to the Milne problem and compute the related 
asymptotic densities. This point, which was only briefly mentioned in Ref. \cite{BN2018} , is fully developed in the Section \ref{S3} of the present paper. 
We obtain in this way an explicit expression of the interpolation coefficient for the specific case we are interested in, that is the case 
of a potential barrier and of purely electron transport. 
The Maxwell-Boltzmann approximation is also discussed in Section \ref{S3}, which is a further simplification that can be introduced in regimes of low densities or high
temperatures.
\par
Finally, in Section \ref{S4} we set up a model of a graphene n-p-n heterojunction and perform some numerical experiments by assuming 
purely electronic transport and sharp potential barrier profile.
We show that our model is able to reproduce, at least in a suitable range of physical parameters, important features that have been highlighted in laboratory experiments.
\section{Quantum transmission conditions}
\label{S2}
In this section we briefly review the transport model across quantum interfaces in graphene, model that has been developed in Ref.\ \cite{BN2018}.
\par
\begin{figure}[h]
\begin{center}
\includegraphics[width=.6\linewidth]{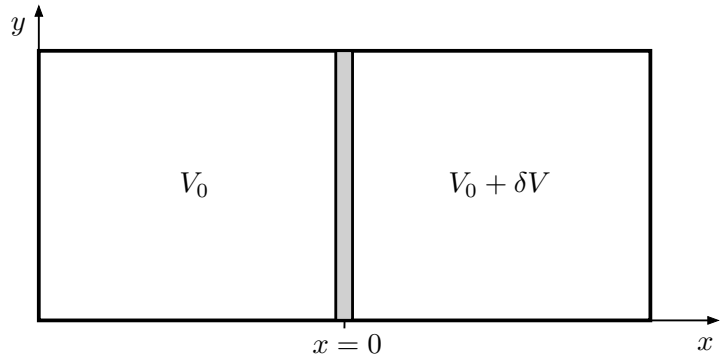}
\caption{Schematic geometry of our model: the rectangle represents the graphene sheet and the central strip represents the quantum active region, i.e.\ the zone where
the variations of $V = V(x)$ are localized. 
Outside the strip, in the two classical regions, the potential $V$ has constant values $V_0$ and $V_0+\deltaV$.}
\label{figura0}
\end{center}
\end{figure}

Assume that a graphene sheet is described by the coordinates $\xx = (x,y)$ and that a ``quantum active region'' (e.g. a potential barrier) is
localized into a tiny strip around $x=0$. 
More precisely, we assume that the electric potential is the sum of two distinct parts, namely
$$
  V(x) + U(x,y),
$$
where $V(x)$ represents the step/barrier profile, which is assumed to have variations localized around $x=0$ and to be constant outside the active region,
taking the values
$$
  V_0 \qquad \text{and} \qquad V_0 + \deltaV 
$$
at the left and at the right, respectively (see Figure \ref{figura0}).
Note that $V_0$ is a ``background'' potential and $\deltaV$ is the total potential variation across the quantum strip.
The ``smooth'' part of the potential,  $U(x,y)$ is assumed to vary on a much larger (macroscopic) space scale 
and can be used to describe, e.g.,  a bias voltage. 
\par
Our the graphene sheet is then modelled as two ``classical'' regions ($x<0$ and $x>0$), where the charge transport is assumed to be diffusive, 
separated by a ``quantum interface'', localized at  $x=0$.
Mathematically, the quantum interface is seen as a boundary where transmission conditions have to be determined by solving a scattering 
problem for the potential $V$.
The derivation of the final macroscopic model  passes firstly through a kinetic step and then a diffusion step, steps briefly described below.
\subsection{Kinetic model}
\label{S2.1}
Let us consider the scattering problem for the electric potential $V$ (the electron potential energy is $-qV$, 
where $q$ is the elementary charge):
\BE
\label{SE}
  \left( -i\hbar \vF  \nabla \cdot \ssg - qV\sigma_0 \right) \psi_{\pp,s} = E\psi_{\pp,s} .
\EE
Here, $\vF$ is the Fermi velocity, $\nabla = (\pt_x,\pt_y)$, $\ssg = (\sigma_1,\sigma_2)$ are the $x$- and $y$- Pauli matrices, 
$\pp = (p_x,p_y)$ is the electron (pseudo)momentum and $E$ is a given energy (which can be either positive or negative).
Moreover, $s = \pm 1$ is the chirality index, denoting electron states of positive and negative chirality \cite{CastroNeto09}.
Solving this equation provides the scattering states $\psi_{\pp,s}$ and the reflection/transmission coefficients $T_s^i(\pp)$,  $R_s^i(\pp)$ 
corresponding to the energy
$$
  E = sv_F\abs{\pp}.
$$
Note that states with positive chirality are also states of positive energy (upper Dirac cone) and states with negative chirality are also
states of negative energy band (Dirac cone).
\par
The upper index $i$ appearing in the coefficients $T_s^i(\pp)$,  $R_s^i(\pp)$ takes the values 1 and 2, and refers to a left ($i=1$) 
or right ($i = 2$) incoming wave. 
Throughout this paper, an upper index $i = 1,2$ will always denote left and right, respectively.
The scattering coefficients satisfy some basic properties:
\begin{enumerate}
\item[\it i)]
$T_s^i(\pp) \geq 0$ and $R_s^i(\pp) \geq 0$, with $T_s^i(\pp) + R_s^i(\pp) = 1$ (unitarity);
\item[\it ii)]
$T_s^i(\pp)$ and $R_s^i(\pp)$ are symmetric with respect to $p_x$ and $p_y$ (symmetry);
\item[\it iii)]
$T_s^1(\pp) = T_{s'}^2(\pp')$ if the conservation of energy 
\BE
\label{CoE}
  s \vF \abs{\pp} = s' \vF \abs{\pp'} - q\,\deltaV,
\EE  
holds (reciprocity).
\end{enumerate}
The key remark is that, away from the quantum interface, the scattering states are superpositions of incoming/reflected/transmitted 
plane-wave-like solutions of the form 
\BE 
\label{planew}
\psi_{\pp,s}(\xx) =
\begin{pmatrix}  1 \\  s\,\e^{i\, \phi} \end{pmatrix} \e^{\frac{i}{\hbar}\pp\cdot\xx},
\EE
which have definite values of chirality and momentum.
Such waves are semi-classically interpreted as inflowing and outflowing particles in the classical regions \cite{NBA98,NBA02}.
More precisely, if the phase-space distributions 
$$
   w^i_s(\xx,\pp), \qquad s = \pm 1, \quad
$$
describe the electron populations with positive energy ($s = +$) and negative energy  ($s = -$) in the two classical regions, $x<0$ ($i = 1$) and $x>0$ (i = 2), we assume that at $x = 0$ the following kinetic transmission condition (KTC) hold:
\BE
\label{KTC}
\left\{ 
\begin{aligned}
&w_s^1(\pp) = R_s^1(\pp) w_s^1( \refl\pp) + T_{s'}^2(\pp') w^2_{s'}(\pp'),&\quad &sp_x,\, s'p'_x < 0,
\\[8pt]
&w_{s'}^2(\pp') = R_{s'}^2(\pp') w_{s'}^2(\refl\pp') + T_s^1(\pp) w^1_s(\pp),& &s'p'_x,\, sp_x > 0,
\end{aligned}
\right.
\EE
where $s$, $s'$, $\pp$ and $\pp'$ satisfy the conservation of energy \eqref{CoE} and the conservation of momentum 
in the $y$ direction:
\BE
\label{pycons}
  p_y = p'_y .
\EE
In \eqref{KTC} we have denoted by a tilde the reflection transformation
\BE
\label{refldef}
    \refl\pp := (-p_x,p_y)
\EE 
and, in order to avoid cumbersome expressions, we have only indicated the dependence on the relevant variable $\pp$, omitting the variables $y$ (which is just a parameter) and
$x$, which is of course equal to $0$ at interface.
The meaning of Eq.\ \eqref{KTC} is clear: the first equation says that the inflow in the left classical region through the interface is partly 
due to reflected particles from the left and  partly due to transmitted particles from the right, and the second equation describes the analogous balance 
of particles inflowing in the right region.
Note that, since negative-chirality electrons travel in the direction opposed to momentum \cite{CastroNeto09}, the conditions $sp_x < 0$ 
and $s'p'_x < 0$ describe leftward particles, while $sp_x > 0$ and $s'p'_x > 0$ describe rightward particles.
\par
Since we shall study the diffusive limit of the kinetic model and, therefore, statistical considerations will come into play, it is convenient 
to switch from positive/negative-energy electrons to electron/holes, by means of the transformation 
\BE
\label{fdef}
  f^i_+(\xx,\pp) = w^i_+(\xx,\pp), \qquad
  f^i_-(\xx,\pp) =  1 - w^i_-(\xx,-\pp).
\EE
Now, $f^i_+$ and $f^i_-$ represent, respectively, the phase-space populations of electrons and holes (both with positive energy).
Note that both electrons and holes move in the same direction of the momentum.
\par
In the classical regions, the dynamics of each population is assumed to be described by the stationary BGK (relaxation time) 
transport equation 
\cite{Barletti14,LF18}
\BE
\label{TE}
  \vv\cdot\nabla_\xx f^i_s  - s q \nabla_\xx U\cdot\nabla_\pp f^i_s = 
\frac{1}{\tau}\left(\feq^i_s - f_s\right)
\EE
where $U$ is the smooth part of the potential, as discussed above, and
\BE
\label{vdef}
   \vv = \vF\,\frac{\pp}{\abs{\pp}} 
\EE
is the semiclassical velocity.
The right-hand side of Eq.\ \eqref{TE} describes the separate relaxation of electrons and holes to the local Fermi-Dirac 
distributions 
\BE
\label{FD}
\feq^i_s(\xx,\pp) = \frac{1}{\e^{\beta \left[\vF \abs{\pp}-A^i_s(\xx)\right]}+1},
\EE
where $\beta = (k_BT)^{-1}$, $T$ being the phonon bath temperature and $k_B$ the Boltzmann constant.
The functions $A^i_s(\xx)$ are defined as
\BE
   A^i_s(\xx) = s q V_0 + \mu^i_s(\xx),
\EE
where $V_0$ is the background potential and  $\mu^i_s$ are the chemical potentials of left and right electrons and holes
(however, in the following we will refer to the functions $A^i_s$ as to ``chemical potentials'').
Since the collisions conserve the number of particles, the chemical potentials are constrained by the relation
\BE
\label{constr}
  n_s^i(\xx) := \bk{f_s^i}(\xx) = \bk{\feq_s^i}(\xx), 
\EE
where
\BE
\label{bkdef}
  \bk{\cdot} = \frac{1}{h^2} \int_{\mR^2} \cdot \, d\pp,
\EE
where $h$ is the Planck constant.
The normalization constant is needed to retrieve the correct spatial density from phase-space density \cite{Barletti14}.
Integration of $\feq^i_s$ yields the following relation between density and chemical potential:
\BE
\label{AvsN}
   \beta A^i_s = \phi_2^{-1} \Big( \frac{n^i_s}{n_0} \Big),
\EE
where 
$$
  \qquad n_0 = \frac{2\pi}{(\beta h \vF)^2}
$$
and 
\BE
\label{phidef}
  \phi_k(z) := \frac{1}{\Gamma(k)} \int_0^\infty \frac{t^{k-1}}{\e^{t-z} + 1}\,dt
\EE
is the Fermi integral of order $k$.
\par
The transport equations \eqref{TE}, which hold separately in $x>0$ ($i=1$) and $x<0$ ($i = 2$), are connected through the quantum interface by assuming that at $x = 0$ 
the KTC \eqref{KTC} holds.\footnote{%
The electron/hole version of \eqref{KTC} is readily obtained by means of the transformation \eqref{fdef}.%
}
It is proven in Ref.\ \cite{BN2018} that the boundary conditions \eqref{KTC} conserve the total charge flux across the interface, namely
\BE
\label{Jcons}
  j^1_{+,x} - j^1_{-,x} = j^2_{+,x} - j^2_{-,x}, \qquad \text{at $x = 0$,}
\EE
where
\BE
\label{Jdef}
 (j^i_{s,x}, j^i_{s,y}) =  \jj^i_s := \bk{\vv f^i_s}
\EE
is the current.
In addition, if $\deltaV = 0$, then the conservation of the flux holds separately for each population 
\BE
\label{Jcons2}
  j^1_{+,x} = j^2_{+,x},  \qquad  j^1_{-,x} = j^2_{-,x},  \qquad \text{at $x = 0$.}
\EE
\subsection{Diffusion model}
\label{S2.2}
The diffusive limit of Eq.\ \eqref{TE} can be obtained by means of the standard machinery of kinetic theory (namely, the Chapman-Enskog expansion) and, in the bulk classical regions, 
yields the following fermionic drift-diffusion equations \cite{JSP12,BN2018} for the surface densities $n^i_s$:
\BE
\label{SDD}
 \DIV \jj^i_s = 0, \qquad \jj^i_s = -\frac{\tau \vF^2}{2}\left[\nabla n^i_s - s  \beta n_0\, \phi_1(\beta A^i_s)  q \nabla U \right],
\EE
where we recall that $A^i_s$ are related to $n^i_s$ by \eqref{AvsN} and $\phi_k$ is given by \eqref{phidef}.
Of course, one can alternatively use the chemical potential as unknown, in which case the drift-diffusion equations take the form
\BE
\label{SDDA}
 \DIV \jj^i_s = 0, \qquad \jj^i_s = -\frac{\pi \tau}{\beta h^2}\,  \phi_1(\beta A^i_s) \nabla \left( A^i_s - s q U \right).
\EE 
Of course our relaxation-time approach is a poor approximation of the electron-phonon scattering and, as a consequence, of the electron mobility. 
This can be at least partially fixed by tuning the parameter $\tau$ at a given temperature.
\par
\smallskip
A more difficult task is to obtain the diffusive limit of the transmission conditions \eqref{KTC}.
This requires a boundary layer analysis, leading to Milne (half-space) kinetic problem. 
The result of such analysis, contained in Ref.\ \cite{BN2018}, can be summarized as follows.
\par
After the introduction of the ``magnified'' boundary-layer variable $\xi = x/\tau$, the analysis 
leads to the introduction of a boundary corrector $\theta_s^i(\xi,y,\pp)$ at order $\tau$ in the Hilbert expansion.
Up to an error of order $\tau^2$ the corrector satisfies the equation
\BE
\label{MilnEq}
 v_x \frac{\pt \theta_s^i}{\pt \xi}  = L^i_s \bk{\theta_s^i} - \theta^i_s,\quad  (-1)^i\xi>0, \quad \pp \in \mR^2,
\EE
where $L^i_s$ is the linearized Fermi-Dirac distribution \eqref{FD} around a given density $n_s^i$, i.e.
\BE
\label{Ldef}
L^i_s  = \frac{d F^i_s}{dn^i_s}   =  \frac{ (F^i_s)^2 \,\e^{\beta ( \vF \abs{\pp} - A^i_s)}}{n_0 \, \phi_1(\beta A^i_s)}.  
\EE
As it is shown in Ref.\ \cite{BN2018}, the four equations \eqref{MilnEq} are coupled at $\xi = 0$ by the following nonhomogeneous version 
of the KTC:
\BE
\label{MilneTC}
\left\{
\begin{aligned}
&\theta_s^1(\pp) - G^1_s(\pp) = R^1_s(\pp)\left[\theta_s^1(\refl\pp) - G_s^1(\refl\pp) \right]  
+ss'  T^2_{s'}(\pp')\left[  \theta^2_{s'}(\pp') - G^2_{s'}(\pp') \right] \ &p_x,\, p'_x < 0,
\\[8pt]
&\theta_{s'}^2(\pp') - G_{s'}^2(\pp')  = R^2_{s'} (\pp)\left[  \theta_{s'}^2(\refl\pp') - G_{s'}^2(\refl\pp') \right] 
  + ss' T_s^1(\pp)\left[ \theta^1_{s}(\pp) - G^1_{s}(\pp)\right],\ &p'_x,\, p_x > 0,
\end{aligned}
\right.
\EE
where 
\BE
\label{Gdef}
 G^i_s(y,\pp) := \frac{2}{\tau \vF^2}\,L^i_s(\xx,\pp)\, \vv(\pp) \cdot \jj^i_s (\xx)_{\mid x = 0}.
\EE
We remark that in Eq.\ \eqref{MilneTC} only the dependence on the relevant variable $\pp$ has been explicitly indicated and, as usual,  $s$, $s'$, $\pp$ and 
$\pp'$ are related by the conservation of energy \eqref{CoE}.
We remark that \eqref{MilnEq}-\eqref{MilneTC} is a system of four Milne (half-space, half-range) problems coupled at $\xi = 0$ by 
nonhomogeneous transmission conditions.
in Ref.\ \cite{BN2018}, the following result is proven, which is a generalization to the multicomponent case of analogous results (obtained, e.g., in 
Refs.\ \cite{BBS1984,DEA02,DS98}).
\begin{theorem}
\label{T1}
For any given $n^i_s \geq 0$ (with $s = \pm1$, $i = 1,2$), problem \eqref{MilnEq}-\eqref{MilneTC} 
admits a solution $(\theta_+^1,\theta_+^2,\theta_-^1,\theta_-^2)$, such that 
$$
\theta_s^i \in \mathrm{L}^\infty\big( (-1)^i[0,+\infty)\times \mR^2,  (L^i_s)^{-1}d\xi d\pp \big),
$$ 
if and only if the flux conservation \eqref{Jcons} (or \eqref{Jcons2}, if $\deltaV = 0$) holds.
This solution is unique up to the addition of any homogeneous solution (i.e., with $G^i_s = 0$).
Moreover, four constants  $n^{i,\infty}_s$ exist s.t.
$$
    \theta_s^i  \to  n_s^{i,\infty} L^i_s \quad \text{{\color{black} as\ } $\xi \to (-1)^i\infty$,}
$$
and the convergence is exponentially fast in $\xi$. 
\end{theorem}
We remark that the four constants $n^i_s$ appear in the definition of $L^i_s$, which is the linearization of the Fermi-Dirac distribution around  $n^i_s$.
We also remark that the coordinate $y$ is an overall parameter in the problem (in particular, $n^i_s$ and  $n^{i,\infty}_s$ may depend on the parameter $y$).
\par
The second main result contained in Ref.\ \cite{BN2018} links the solution to the Milne problem \eqref{MilnEq}-\eqref{MilneTC} 
with the diffusion limit at the interface.
%
%
\begin{theorem}
\label{T2}
Let $n^1_s$ and $n^2_s$ be the left and right densities at $x=0$ and let 
$$
  A[n] =  \frac{1}{\beta}\,\phi_2^{-1} \left( \frac{n}{n_0} \right)
$$ 
denote the chemical potential $A$ corresponding to the density $n$, according to \eqref{AvsN}.
Then, up to $\cO(\tau^2)$, the condition
\BE
\label{DTC}
sA[n^1_s + \tau n_s^{1,\infty}] = s'A[n^2_{s'} + \tau n_{s'}^{2,\infty}]  - q \,\deltaV,
\EE
hold for all couples $(s,s')$ satisfying the conservation of energy \eqref{CoE} for some $\pp$ and $\pp'$ in a nonzero measure set,
where $n_{s}^{i,\infty}$ are the asymptotic densities of the solution to the Milne problem \eqref{MilnEq}-\eqref{MilneTC} (see Theorem \ref{T1}).
Moreover,  condition \eqref{DTC} is not affected by the particular choice of the solution to \eqref{MilnEq}-\eqref{MilneTC}.
\end{theorem}
By expanding Eq.\ \eqref{DTC} at first order in $\tau$, and using the property $\phi_k' = \phi_{k-1}$ of Fermi functions, 
we obtain another version of Eq.\ \eqref{DTC}:
\BE
\label{DTC1}
sA^1_s - s'A^2_{s'} + q\,\deltaV = \frac{\tau}{\beta n_0} \left( s'\alpha^2_{s'} n_{s'}^{2,\infty} -  s\alpha^1_s n_s^{1,\infty} \right)
\EE
where 
\BE
\label{alphadef}
\alpha^i_s := \frac{1}{\phi_1(\beta A^i_s)},
\EE
which is a more explicit condition on the left and right chemical potentials $A^i_s = A[n^i_s]$.
Equation  \eqref{DTC1} gives the diffusive transmission conditions (DTC) that connect the two classical regions at the two sides of $x=0$.
They contain the quantum information coming from the scattering problem \eqref{SE}, which is enclosed in the four asymptotic densities $n_{s}^{i,\infty}$.
Note that at leading order in $\tau$ we obtain the semiclassical condition  
\BE
\label{DTC0}
sA^1_s - s'A^2_{s'} = - q\,\deltaV,
\EE
in which case the quantum dynamics occurring at the interface is completely lost.
\par

\section{Evaluation of the asymptotic densities}
\label{S3}
Solving the Milne problem \eqref{MilnEq}-\eqref{MilneTC}, which is needed in order to obtain the asymptotic densities $n_{s}^{i,\infty}$, 
implies that a  ``kinetic'' stage is still present in our diffusive model. 
This is not very appealing, when looking for a simple and numerically treatable model.  
Then, we should resort to some kind of approximation of the solution of the Milne problem.
\subsection{Albedo approximation}
\label{S3.1}
A typical approach \cite{DEA02,DS98} consists in finding some approximation of the ``albedo operator'', i.e., the map that connects the inflow, 
$\theta^i_s(0,y,\pp)$, $(-1)^i p_x > 0$, to the outflow $\theta^i_s(0,y,\pp)$, $(-1)^i p_x < 0$, where  $\theta^i_s(\xi,y,\pp)$ is a solution
to Eq.\ \eqref{MilnEq}.
In particular, assuming that the collisions are very fast, one can look for an approximate outflow of the equilibrium form
\BE
\label{AA1}
   \theta^i_s(0,y,\pp) = L^i_s(y,\pp) \rho^i_s(y), \qquad (-1)^i p_x < 0,
\EE
where $\rho^i_s(y)$ are outflow densities subject to the constraint of vanishing flux at $\xi = 0$:
\BE
\label{AA2}
     \int_{\mR^2} \theta^i_s(0,y,\pp) \,v_x(\pp)\, d\pp = 0.
\EE
To avoid cumbersome notations, in the following we will omit the explicit indication of the variables $\xi = 0$, $y$ and $p_y$, when not necessary.
By using the properties of the scattering coefficients we can rewrite the first of equations \eqref{MilneTC}  in the following way:
\begin{multline}
\label{MilneTC2}
\theta_s^1(\pp) - \theta_s^1(\refl\pp)  + sT^1_s(\pp)\left[s\theta_s^1(\refl\pp) - s'\theta^2_{s'}(\pp')\right]
\\
= G_s^1(\pp) - G_s^1(\refl\pp)  + sT^1_s(\pp)\left[sG_s^1(\refl\pp) - s'G^2_{s'}(\pp')\right], \qquad p_x,\, p'_x < 0,
\end{multline}
Let us multiply this equation by $v_x$  and integrate over the inflow range $\{ \pp \in \mR^2 \mid p_x<0\}$  (simply denoted by ``$p_x<0$''). 
This yields
\begin{multline}
\label{AUX1}
\int \theta_s^1(\pp)\, v_x\, d\pp  + \int_{p_x<0} sT^1_s(\pp)\left[s\theta_s^1(\refl\pp) - s'\theta^2_{s'}(\pp')\right] v_x \,d\pp
\\
= \int G_s^1(\pp)\, v_x\,d\pp +  \int_{p_x<0} sT^1_s(\pp)\left[sG_s^1(\refl\pp) - s'G^2_{s'}(\pp')\right] v_x \,d\pp, 
\end{multline}
We now recall that $ \theta_s^i$ is approximated by \eqref{AA1} and that the null-flux condition \eqref{AA2} holds. 
We recall, moreover,  that $G_s^i$ is given by \eqref{Gdef}. 
Then, 
\begin{multline}
\label{AUX2}
\int_{p_x<0} s T^1_s(\pp)\left[sL_s^1(\pp)\rho^1_s - s'L^2_{s'}(\pp')\rho^2_{s'}\right] v_x \,d\pp =
 \jj^1_s \cdot \frac{2}{\tau\vF^2} \int  L^1_s(\pp) \vv v_x\,d\pp 
\\
 +  \frac{2}{\tau\vF^2} \int_{p_x<0} sT^1_s(\pp)\left[sL_s^1(\pp)\refl\vv\cdot\jj^1_s - s'L^2_{s'}(\pp')\vv'\cdot\jj^2_{s'}\right] v_x \,d\pp, 
\end{multline}
We shall now use the identity 
\BE
\label{aLLa}
  \alpha^2_{s'}L^1_s(\pp) = \alpha^1_s L^2_{s'}(\pp'),
\EE
which holds assuming $sA^1_s - s'A^2_{s'} = -q\,\deltaV$, i.e., at leading order in $\tau$  (see Ref.\ \cite{BN2018}). 
Using this relation to compute $n^{i,\infty}_s$ will produce an error of order two in \eqref{DTC1}.
In Ref.\ \cite{BN2018} it is also proven that
\BE
\label{Lvar}
   \int L^i_s(\pp) \vv \otimes\vv\,d\pp = \frac{h^2 \vF^2}{2} I .
\EE
From \eqref{AUX2}, \eqref{aLLa} and \eqref{alphadef}, we obtain
\begin{multline}
\label{SY1}
\int_{p_x>0} \frac{sT^1_s(\pp)L_s^1(\pp)}{\alpha^1_s} X_{ss'} v_x \,d\pp =    \frac{h^2}{\tau}\,j^1_{s,x}  
\\
   -   \frac{2}{\tau\vF^2} \int_{p_x>0}  \frac{sT^1_s(\pp)L_s^1(\pp)}{\alpha^1_s} 
  \left[s\alpha^1_s (v_x)^2 j^1_{s,x} +   s'\alpha^2_{s'} v'_x v_x  j^2_{s',x}\right] d\pp, 
\end{multline}
where 
 \BE
\label{Xdef}
X_{ss'} := s'\alpha^2_{s'} \rho^2_{s'} - s\alpha^1_s \rho^1_s \, .
\EE
and the signs where chosen such that both $v_x$ and $v'_x$ are positive.
\par
With an analogous procedure, from the second of equations \eqref{MilneTC} we obtain
\begin{multline}
\label{SY2}
\int_{p'_x>0} \frac{s'T^2_{s'}(\pp')L_{s'}^2(\pp')}{\alpha^2_{s'}} X_{ss'} v'_x \,d\pp =  \frac{h^2}{\tau}\,j^2_{s',x}  
\\
 +   \frac{2}{\tau\vF^2} \int_{p'_x>0}  \frac{s'T^2_{s'}(\pp')L_{s'}^2(\pp')}{\alpha^2_{s'}} 
  \left[s'\alpha^2_{s'} (v'_x)^2 j^2_{s',x} +  s\alpha^1_s v_x v'_x  j^1_{s,x}\right] d\pp', 
\end{multline}
where, again, signs have been chosen such that both $v_x$ and $v'_x$ are positive.
Note that in \eqref{SY1} $s'$ and $\pp'$ depend on $\pp$ and $s$ while, conversely,  in  \eqref{SY2} $ss$ and $\pp$ depend on $\pp'$ 
and $s'$ (through the conservation of energy \eqref{CoE}).
However, once the integrals are split in the zones where $s'$ and, respectively, $s$ are constant, equations \eqref{SY1} and \eqref{SY2}
become a linear system for the unknowns $X_{ss'}$.
For example, assuming $\deltaV < 0$ one obtains
\BE
\label{SY}
\left\{
\begin{aligned}
  &D_{++} X_{++} + D_{+-} X_{+-} &= H^1_+
  \\
  -&D_{--} X_{--} &= H^1_- 
  \\
  &D_{++} X_{++}  &= H^2_+
  \\
  -&D_{--} X_{--} - D_{+-} X_{+-} &= H^2_- 
  \end{aligned}
\right.
\EE
where
$$
\begin{aligned}
&D_{++}  = \int\limits_{p_x > 0 \atop \vF\abs{\pp} > \abs{\deltaV}} \frac{T^1_+ (\pp) L^1_+ (\pp)}{\alpha^1_+} \,v_x \,d\pp
   =  \int\limits_{p'_x > 0} \frac{T^2_+ (\pp') L^2_+ (\pp')}{\alpha^2_+} \,v'_x \,d\pp',
\\[4pt]
&D_{+-}  = \int\limits_{p_x > 0 \atop  \vF\abs{\pp} < \abs{\deltaV}} \frac{T^1_+ (\pp) L^1_+ (\pp)}{\alpha^1_+}\,v_x \,d\pp
  =  \int\limits_{p'_x > 0 \atop \vF\abs{\pp'} < \abs{\deltaV}} \frac{T^2_+ (\pp') L^2_+ (\pp')}{\alpha^2_+} \,v'_x \,d\pp',
\\[4pt]
&D_{--}  = \int\limits_{p_x > 0} \frac{T^1_- (\pp) L^1_- (\pp)}{\alpha^1_-}\,v_x \,d\pp
     =  \int\limits_{p'_x > 0 \atop \vF\abs{\pp'} > \abs{\deltaV}} \frac{T^2_- (\pp') L^2_- (\pp')}{\alpha^2_-} \,v'_x \,d\pp',
\\[4pt]
&H^1_s = \frac{h^2}{\tau}\,j^1_{s,x}   -   \frac{2}{\tau\vF^2}\int_{p_x>0}  \frac{T^1_s(\pp)L_s^1(\pp)}{\alpha^1_s} 
  \left[\alpha^1_s v_x^2 j^1_{s,x} +   ss'\alpha^2_{s'} v'_x v_x  j^2_{s',x}\right] d\pp,
\\[4pt]
&H^2_{s'} = \frac{h^2}{\tau}\,j^2_{s',x}  +   \frac{2}{\tau\vF^2} \int_{p'_x>0}  \frac{T^2_{s'}(\pp')L_{s'}^2(\pp')}{\alpha^2_{s'}} 
  \left[\alpha^2_{s'} {v'_x}^2 j^2_{s',x} +  ss'\alpha^1_s v_x v'_x  j^1_{s,x}\right] d\pp'.
\end{aligned}
$$
Note that the four equations in \eqref{SY} are not independent because the difference of the first two is equal to the difference of
the second two, which can be verified directly by using $v_xd\pp = v'_xd\pp'$ and the flux conservation $j^1_{+,x} - j^1_{-,x} = j^2_{+,x} - j^2_{-,x}$ 
(or derived from general considerations on the structure of KTC, see the proof of Proposition 3.1 in Ref.\ \cite{BN2018}).
Hence, \eqref{SY} is a rank-3 system for the unknowns $X_{++}$, $X_{+-}$ and $X_{--}$. 
The case $\deltaV = 0$ will be examined in the next subsection.
\par
System \eqref{SY} allows to compute the $X_{ss'}$'s as functions of the currents $j^1_{s,x}$ and the densities $n^1_s$ at the interface
(the latter are ``hidden'' in the terms $L^i_s$ and $\alpha^i_s$).
We now need to relate the asymptotic densities $n^{i,\infty}_s$ to the quantities $X_{ss'}$.
In order to do this, let us consider any function $\theta_s^i$ that satisfies the half-space equation \eqref{MilnEq}.
Integrating in $\pp$ yields
$$
 \frac{\pt \bk{v_x\theta_s^i}}{\pt \xi}  = 0, 
$$
which implies that the current is constant. 
Using the fact that  $\theta_s^i  \to  n_s^{i,\infty} L^i_s$ as  $\xi \to (-1)^i\infty$ (see Theorem \ref{T1})  we obtain that such constant is zero:
$$
  \bk{v_x\theta_s^i} = 0.
$$
Then, multiplying Eq.\ \eqref{MilnEq} by $v_x$ and integrating in $\pp$ yields
$$
  \frac{\pt \bk{v_x^2\theta_s^i}}{\pt \xi}  = 0, 
$$
which means that any solution to \eqref{MilnEq} has constant\footnote{Possibly depending on $y$.} variance.
To evaluate this constant we use again the asymptotics $\theta_s^i  \to  n_s^{i,\infty} L^i_s$ and the identity \eqref{Lvar},  
and finally obtain
\BE
\label{asyvar}
  \bk{v^2_x\theta_s^i}  = \frac{1}{h^2} \int v^2_x\theta_s^i\, d\pp=  \frac{\vF^2}{2}\,n_s^{i,\infty}.
\EE
Then, let us multiply by $v_x^2$ the first of the two equations \eqref{MilneTC}, written in the form  \eqref{MilneTC2}, 
and integrate with respect to $\pp$ over the inflow range $p_x<0$.
When doing so, note that
$$
\int_{p_x<0} \left[ \theta_s^1(\pp) - \theta_s^1(\refl\pp) \right] v_x^2 \,d\pp = 
\int  \theta_s^1(\pp)\, v_x^2 \,d\pp - 2\rho^i_s \int_{p_x>0} L^i_s(\pp) \, v_x^2 \,d\pp
$$ $$
  \int  \theta_s^1(\pp)\, v_x^2 \,d\pp - \rho^i_s \int L^i_s(\pp) \, v_x^2 \,d\pp = 
  \frac{h^2 \vF^2}{2}\left( n_s^{i,\infty} - \rho^i_s \right).
$$
We obtain in this way
\begin{multline}
\label{SYaux1}
\frac{h^2 \vF^2}{2}\left( n_s^{1,\infty} - \rho^1_s \right) - \int_{p_x>0} \frac{sT^1_s(\pp)L_s^1(\pp)}{\alpha^1_s} X_{ss'}\,v_x^2 \,d\pp =    
-\frac{4}{\tau \vF^2}\,j^1_{s,x}  \int_{p_x>0} L_s^1(\pp)\,v_x^3 \, d\pp
\\
   + \frac{2}{\tau \vF^2} \int_{p_x>0}  \frac{sT^1_s(\pp)L_s^1(\pp)}{\alpha^1_s} 
  \left[s\alpha^1_s v_x^3 j^1_{s,x} +   s'\alpha^2_{s'} v'_x v_x^2  j^2_{s',x}\right] d\pp
\end{multline}
(with positive $v'_x$). 
Analogously,
\begin{multline}
\label{SYaux2}
\frac{h^2 \vF^2}{2}\left( n_{s'}^{2,\infty} - \rho^2_{s'} \right) + \int_{p'_x>0} \frac{s'T^2_{s'}(\pp')L^2_{s'}(\pp')}{\alpha^2_{s'}} X_{ss'}\,{v'_x}^2 \,d\pp' =    
\frac{4}{\tau \vF^2}\,j^2_{s',x}  \int_{p'_x>0} L^2_{s'}(\pp')\,{v'_x}^3\, d\pp'
\\
   -  \frac{2}{\tau \vF^2}\int_{p'_x>0}  \frac{s'T^2_{s'}(\pp')L^2_{s'}(\pp)}{\alpha^2_{s'}} 
  \left[s'\alpha^2_{s'} {v'_x}^3 j^2_{s',x} +   s\alpha^1_s v_x {v'_x}^2  j^1_{s,x}\right] d\pp'
\end{multline}
(with positive $v_x$). 
Multiplying \eqref{SYaux1} by $\frac{2s\alpha^1_s}{h^2\vF^2}$ and \eqref{SYaux2} by $\frac{2s'\alpha^2_{s'}}{h^2\vF^2}$, and subtracting the former from the latter, we finally obtain
\begin{multline}
\label{Albedo1}
  s'\alpha^2_{s'}  n_{s'}^{2,\infty} -  s\alpha^1_s  n_s^{2,\infty}
  =  X_{ss'} 
  - \frac{2}{h^2\vF^2}  \int_{p_x>0}T^1_s(\pp)L_s^1(\pp) X_{ss'(\pp)}\,v_x^2 \,d\pp
 \\
  - \frac{2}{h^2\vF^2} \int_{p'_x>0}T^2_{s'}(\pp')L^2_{s'}(\pp') X_{s(\pp')s'}\,{v'_x}^2 \,d\pp' 
   + E_{ss'},
\end{multline}
where
\begin{multline}
\label{Albedo2}
 E_{ss'} = \frac{8s\alpha^1_s j^1_{s,x}}{\tau h^2\vF^4}   \int_{p_x>0} L_s^1(\pp)\,v_x^3 \, d\pp
  + \frac{8s' \alpha^2_{s'} j^2_{s',x}}{\tau h^2\vF^4}   \int_{p'_x>0} L^2_{s'}(\pp')\,{v'_x}^3\, d\pp'
 \\
    - \frac{4}{\tau h^2 \vF^4}  \int_{p_x>0}  T^1_s(\pp)L_s^1(\pp)
  \left[s\alpha^1_s v_x^3 j^1_{s,x} +   s'\alpha^2_{s'(\pp)} v'_x v_x^2  j^2_{s'(\pp),x}\right] d\pp
 \\
  - \frac{4}{\tau h^2  \vF^4} \int_{p'_x>0}  T^2_{s'}(\pp')L^2_{s'}(\pp)
  \left[s'\alpha^2_{s'} {v'_x}^3 j^2_{s',x} +   s\alpha^1_{s(\pp')} v_x {v'_x}^2  j^1_{s(\pp'),x}\right] d\pp'.
\end{multline}
Note that in the integrals above, in order to avoid possible confusion, we explicitly denoted the dependence of $s'$ on $\pp$ and of $s$ on $\pp'$.
\par
Expressions \eqref{Albedo1} and  \eqref{Albedo2} give the term  $s'\alpha^2_{s'}  n_{s'}^{2,\infty} -  s\alpha^1_s  n_s^{2,\infty}$ as a function of $j^i_s$, $n^i_s$ and $X_{ss'}$, the latter being given by system \eqref{SY}.
This is exactly the term occurring in the DTC \eqref{DTC1}.
\begin{remark} \rm
A simplified approach, sometimes referred to as the Marshack approximation, consists in stopping the above procedure at the level of Eq.\ \eqref{SY}, and 
taking
$$
   s'\alpha^2_{s'}  n_{s'}^{2,\infty} -  s\alpha^1_s  n_s^{2,\infty} \approx X_{ss'} \ .
$$
This, of course amounts to approximating $\rho^i_s \approx n_s^{i,\infty}$, which in turn means that the collisions are assumed to be so fast that the distribution attains the asymptotic state very close to $x=0$. 
The Marshak approximation can also be considered the first step of a systematic iteration procedure proposed by Golse and Klar \cite{GK95}.
However, numerical experiments suggest that the Marshack approximation can be a very reliable alternative to the more complex 
approaches \cite{DEA02}.
\end{remark}
\subsection{The case $\deltaV = 0$}
\label{S3.2}
If $\deltaV = 0$, then we only have the cases $s = s' = +1$ and $s = s' = -1$. 
Then, the DTC \eqref{DTC} hold in the form $A[n^1_s + \tau n_s^{1,\infty}] =  A[n^2_s + \tau n_s^{2,\infty}]$, $s = \pm 1$, 
which simply implies 
\BE
\label{DTCV0}
  n^1_s -n^2_s = \tau \left( n_s^{2,\infty} -  n_s^{1,\infty} \right), \qquad s = \pm 1.
\EE
Note that electrons and holes are completely decoupled by such DTC. 
Of course, the electron-hole coupling is still present in the Schr\"odinger equation \eqref{SE} and strongly affects 
the scattering coefficients.
\par
When applying the albedo approximation to this case,  we have that $D^1_{+-} = 0$ and $H^i_+ = H^i_-$, $i = 1,2$. 
Hence, \eqref{SY} is a rank-2 system for the unknowns $X_{++}$ and $X_{--}$, where the first two equations are decoupled 
and are equivalent to the second two. 
This is readily solved and yields, explicitly,
\BE
\label{XdV0}
 X_{ss} = \frac{2s \alpha^1_s j^1_{s,x} }{\vF} \, \frac{ \displaystyle \frac{h^2}{2} -  2\int_{p_x>0}  T^1_{s}(\pp)L^1_s(\pp) \Big(\frac{p_x}{\abs{\pp}}\Big)^2 d\pp }
 {\displaystyle  \int_{p_x>0}  T^1_{s}(\pp)L^1_s(\pp) \frac{p_x}{\abs{\pp}}\,d\pp }
\EE
where we used the fact that $\pp' = \pp$, $\jj^1_s = \jj^2_s$ (see \eqref{Jcons2}) and the fact that, at leading order, $A^1_s = A^2_s$, 
which in turn implies $\alpha^1_s = \alpha^2_s$.
Note that the above expression is written for the upper index 1 but it is actually equal to the same expression written for the upper index 2.
In the Marshak approximation, \eqref{XdV0} is already the expression for $\alpha^2_s n_s^{2,\infty} -  \alpha^1_s n_s^{1,\infty}$ and,
comparing with \eqref{DTCV0}, we obtain therefore
\BE
\label{MarshakDTC}
  n^1_s - n^2_s =   s j_{s,x} \vartheta_s, 
  \qquad s = \pm 1.
\EE
where $j_{s,x} := j^1_{s,x} = j^2_{s,x}$ is the common value of the left and right $x$-current, and
\BE
\label{qFD}
  \vartheta_s := \frac{2}{\vF} \, 
  \frac{ \displaystyle \frac{h^2}{2}   -  2\int_{p_x>0}  T^1_{s}(\pp)L^1_s(\pp) \Big(\frac{p_x}{\abs{\pp}}\Big)^2 d\pp }
 {\displaystyle\int_{p_x>0}  T^1_{s}(\pp)L^1_s(\pp) \frac{p_x}{\abs{\pp}}\,d\pp } .
\EE
The quantity $\vartheta_s$, having the dimensions of an inverse velocity, is the analogous of the ``extrapolation coefficient''  typically arising from kinetic boundary layers
 \cite{BBS1984,DEA02,DS98}. 
 Since, in our case, the boundary layer connects two regions, we shall rather call our $q_s$ an ''interpolation coefficient''.
Let us also remark that the disappearance of $\tau$ in Eq.\ \eqref{MarshakDTC} is not contradictory, since the current $ j^1_{s,x}$ already contains the factor $\tau$  (see
Eq.\  \eqref{SDD}).
\par
The reformulation of  the complete albedo approximation \eqref{Albedo1}-\eqref{Albedo2} in the case $\deltaV = 0$ 
is just matter of straightforward computations which are not worth to be reported here.
Instead, it can be interesting to examine the form of the DTC in the Maxwell-Boltzmann (M-B) limit, i.e., for high temperatures or low carrier densities.
We first note that such limit is relevant only in the case of negligible $\deltaV$. 
This can be readily seen by considering the leading-order conditions  \eqref{DTC0}, which for (e.g.)  $\deltaV < 0$ are 
$$
  A^1_+ = A^2_+ - q\,\deltaV, \qquad A^1_+ = - A^2_- - q\,\deltaV, \qquad A^1_- = A^2_- + q\,\deltaV.
$$
These three conditions are clearly incompatible with the requirement  $A^i_s < 0$, ($s = \pm 1$, $i = 1,2$) which is needed for the 
M-B approximation to be valid for both populations on both sides.
The same problem does not arise in the case $\deltaV = 0$, since the leading-order conditions are
$$
  A^1_+ = A^2_+    \qquad   A^1_- = A^2_-,
$$
which are compatible with the M-B regime.
\par
From the mathematical point of view the M-B asymptotic regime  corresponds to the limit $\beta A^i_s \to -\infty$.
In this limit we have 
$$
    \feq^i_s(\xx,\pp) \approx \e^{-\beta \left[\vF \abs{\pp}-A^i_s(\xx)\right]}
    \qquad \text{and} \qquad
    \phi_k(\beta A^i_s) \approx \e^{\beta A^i_s} \approx \frac{n^i_s}{n_0}
$$
(independently on $k$), so that, 
\BE
\label{MBlimit}
  \feq^i_s(\xx,\pp) \approx n^i_s(\xx) M(\pp)
  \qquad \text{and} \qquad
   L^i_s (\xx,\pp) \approx M(\pp)
\EE
where 
\BE
  M(\pp) := \frac{1}{n_0}\,\e^{-\beta\vF \abs{\pp}}
\EE
is the ``Maxwellian'' normalized with respect to $\bk{\cdot}$. 
Then, the M-B approximations of the drift-diffusion equations \eqref{SDD} and \eqref{SDDA} are, respectively,
\BE
\label{SDDM}
 \DIV \jj^i_s = 0, \qquad \jj^i_s = -\frac{\tau \vF^2}{2}\left(\nabla n^i_s - s \beta  n^i_s q \nabla U \right)
\EE
and
\BE
\label{SDDAM}
 \DIV \jj^i_s = 0, \qquad \jj^i_s = -\frac{\pi \tau}{\beta h^2}\,  \e^{\beta A^i_s} \nabla \left( A^i_s - s q U \right),
\EE 
while the form of the DTC remains \eqref{DTCV0}.
Moreover, in the Marshak approximation, the DTC still take the form \eqref{MarshakDTC}, but the
interpolation coefficient is now given by 
\BE
\label{qMB}
  \vartheta_s  =   \frac{2}{\vF} \, 
 \frac{ \displaystyle \frac{h^2}{2}   -  2\int_{p_x>0}  T^1_{s}(\pp) M(\pp) \Big(\frac{p_x}{\abs{\pp}}\Big)^2 d\pp }
 {\displaystyle\int_{p_x>0}  T^1_{s}(\pp) M(\pp) \frac{p_x}{\abs{\pp}}\,d\pp } .
\EE
\par
A simple way to understand the physical meaning of the interpolation constant is the following.
From \eqref{MarshakDTC} and \eqref{AvsN} we can write
$$
   n_0 \phi_2(\beta A^1_s) - n_0 \phi_2(\beta A^2_s) =  s j_{s,x} \vartheta_s .
$$
Then, assuming that $A^i_s$ is not too far from the background potential $seV_0$, we can approximate  
$$
  \phi_2(\beta A^1_s) - \phi_2(\beta A^2_s) \approx  \beta \phi_1(s \beta e V_0)  ( A^1_s - A^2_s )
$$
(recall that $\phi_1$ is the derivative of $\phi_2$), obtaining therefore 
$$
   \frac{A^1_s - A^2_s}{j_{s,x}} \approx \frac{s  \vartheta_s}{\beta  \phi_1(s \beta eV_0)\, n_0}.
$$
We see therefore that  $\vartheta_s$ is proportional to the ratio between the potential variation across the barrier
and the current, and so it is a quantity related to the``quantum resistance''  of the barrier.
Let us finally remark that, for small transmission coefficients, the denominator in expression \eqref{qFD} dominates, yielding a Landauer-like (but
corrected with statistics) expression of the conductance \cite{CF2006}. 
In general, however, the variance-like term  at the numerator of \eqref{qFD} is not negligible. 
Such term can be interpreted as a diffusion correction to the Landauer (ballistic) picture.
\section{Device modelling}
\label{S4}
\begin{figure}
\begin{center}
\includegraphics[width=.6\linewidth]{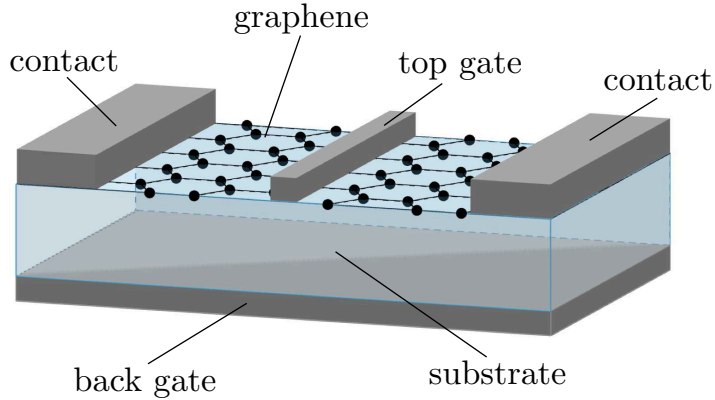}
\caption{A schematic picture of a n-p-n graphene device:  
the graphene sheet is represented as the black honeycomb (not in scale), the grey regions represent gates and contacts, 
and the blue box represents some substrate layer (typically an oxide).}
\label{figura1}
\end{center}
\end{figure}

We give an example of application of the above-developed theory to the modelling of a graphene device.
The architecture that we have in mind is that of a ``n-p-n graphene heterojunction'', which is of primary importance for theoretical investigations 
as well as for possible technological applications \cite{Huard2007,Osyilmaz2007,YK2009}.
In such devices, a relatively thin potential barrier (the  p-region) is obtained as the combined effect of the electrostatic potentials 
of a local top gate ({\it tg})  and a background gate ({\it bg}); additional gates correspond to contacts where a potential bias is applied 
(see the schematic device pictured in Figure \ref{figura1}). 
Let us remark that, since $V$ is a potential barrier, then $\deltaV = 0$.
We shall make the following assumptions:
\begin{enumerate}
\item
the system is homogeneous in the $y$-direction;
\item
the background potential is high enough so that only the population of electrons contribute significantly to the current;
\item
the electron statistics can be described in the Maxwell-Boltzmann approximation.
\end{enumerate}
These assumptions imply that the drift-diffusion equation for electrons in the classical regions (i.e.\ the n-regions) takes the form 
\BE
\label{DD}
 -\frac{\tau \vF^2}{2}\left(\pt_x n^i + \beta  n^i \cE \right) = j,
\EE
where $j$ is the $x$-component of the current (which is constant all along the device) and, assuming that $U$ 
is the linear potential determined by the applied bias,  $\cE = -q\, \pt_x U$ is the corresponding constant electric force.
Note that we have dropped the index $s$, which is equal to $+1$, since holes can be neglected.
\par
Assuming the device length to be $2L$, and the barrier to be located at $x =0$, we have to set boundary conditions at $x = \pm L$ 
and transmission conditions at $x = 0$. 
The boundary conditions are the usual Dirichlet type conditions
\BE
\label{BC}
  n^1(-L) = n_l, \qquad  n^1(L) = n_r,
\EE
where $n_l$ and $n_r$ are the electron densities at the left and right contacts, while for the transmission condition we use the Marshak form 
\eqref{MarshakDTC}, which, in the simplified notation introduced in this section, reads as follows:
\BE
\label{STC}
  n^1- n^2 =   j \vartheta, 
\EE
where $\vartheta \equiv \vartheta_+$ is the interpolation coefficient given by Eq.\ \eqref{qMB}.
\par
As a final ingredient we need to choose a model for the barrier and the transmission coefficient.
We assume a, perfectly sharp and flat, rectangular barrier of width $D$ and energy height 
\BE
\label{Ehdef}
 E_h = - qc_{\mathit{bg}}  V_{\mathit{bg}}  -q c_{\mathit{tg}} V_{\mathit{tg}},
\EE
where $V_{\mathit{bg}}$, $V_{\mathit{tg}}$ are the back gate and local gate voltages, and $c_{\mathit{bg}}$, $c_{\mathit{tg}}$,
are suitable constants that relate the gate voltages to the effective electric potential on the graphene surface
(so that the background potential $V_0$ introduced in Sect.\ \ref{S1} is given by $V_0 = c_{\mathit{bg}} V_{\mathit{bg}}$).
Such constants furnish a simplified description of the (more complicated) capacitive coupling between gates, substrates and graphene,
which is widely used in literature, see e.g.\ Refs. \cite{Fang07,TSS2010,YK2009}.
\par
For a such perfectly sharp barrier the transmission coefficient is given by \cite{CastroNeto09,KNG2006}
\BE
\label{Texpr}
 T_s(\pp) =  \Re\left\{\frac{\cos^2\phi \, \cos^2\phi^*} 
  {[\cos(Dq_x)\, \cos\phi\, \cos\phi^*]^2 + \sin^2(Dq_x) (1 - ss'\sin\phi\, \sin\phi^*)}
  \right\}.
\EE
Here, $\phi$ is the incidence angle, so that $(p_x,p_y) = \abs{\pp} (\cos \phi, \sin\phi)$, 
$s' = \mathop{\mathrm{sgn}}(s\vF\abs{\pp} - E_h)$
is the chirality of the electron inside the barrier,
$$
  q_x = \sqrt{ \left( \frac{s\vF\abs{\pp} - E_h}{\hbar\vF} \right)^2 - 
  \left( \frac{p_y}{\hbar}\right)^2}
$$
is the $x$-component of the refracted momentum inside the barrier and
$$
  \phi^*  = \tan^{-1} \left(\frac{p_y}{\hbar q_x}\right).
$$
is the refraction angle.
Note that $T_s(\pp)$ is independent on the side-index $i$, for obvious symmetry reasons.
The transmission coefficient \eqref{Texpr} (with $s = +1$) will be used in the expression \eqref{qMB} for $q$. 
\par
\begin{figure}
\begin{center}
\includegraphics[width=.9\linewidth]{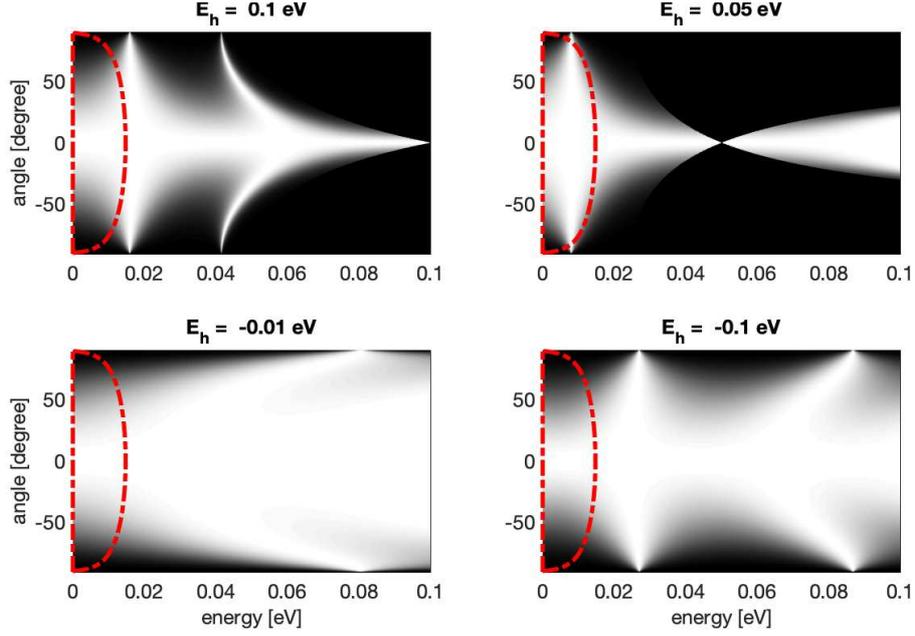}   
\caption{Gray-scale plots of $T_+(\pp)$, as a function of the energy $E = \vF\abs{\pp}$ and of the incidence angle $\phi$, for
different values of the energy height $E_h$. White corresponds to perfect transmission ($T_+ = 1$) and black to total reflection ($T_+ = 0$).
Note that for $\phi = 0$ the barrier is always completely transparent, regardless to $E_h$,  which is the 
so-called Klein paradox \cite{KNG2006}.
The dashed red line is a contour line of $M(\pp) \cos\phi$, corresponding to a region that encompasses approximately 
90\% of its integral;  such region is therefore where the main contribution to the integrals in \eqref{qMB} comes from 
(the same region for $M(\pp) \cos^2\phi$ is just slightly narrower). In this figure the barrier width is $50\,\mathrm{nm}$ 
and the temperature is $40\,\mathrm{K}$. For lower values of the temperature, the Maxwellian will be narrower, resulting in a higher
sensitivity to the variations of $T_+$.}
\label{figura2}
\end{center}
\end{figure}

In Figure \ref{figura2} we represent $T_+(\pp)$, as a function of the energy $E = \vF\abs{\pp}$ and the incidence angle $\phi$, for
different values of the energy height $E_h$ of the barrier, together with the region which is significant for the integrals in \eqref{qMB}.
It is evident from the figure that changing the value of $E_h$ produces significant variations of the integrals, resulting in variations 
of the interpolation coefficient. 
\subsection{Numerical results}
\label{S4.1}
In order to numerically solve the problem \eqref{DD}-\eqref{BC}-\eqref{STC} we adopt a simple finite-difference scheme that can be
outlined as follows.
Each of the two spatial domains, $[-L,0]$ and $[0,L]$, is decomposed in $N$ cells of length $\Delta x$, labeled with an index $k$, increasing 
in the $x$-direction.
The corresponding discretized values of the density are $n^1, n^2, \ldots, n^N$ and in the left region and  
$n^{N+1}, n^{N+2}, \ldots, n^{2N}$ in the right region.
The drift-diffusion equation for the is therefore discretized as
\BE
\frac{n^{k-1}-2n^{k}+n^{k+1}}{\Delta x^2} - \beta \cE \frac{n^{k+1}-n^{k-1}}{2\Delta x}=0,
\EE
with $k = 2, \ldots, N-1$, and $k = N+1, \ldots, 2N-1$. 
At $x=\pm L$ we impose the Dirichlet boundary conditions \eqref{BC}
\BE
n^1 = n_l, \qquad n^{2N}=n_r.
\EE
At the interface $x=0$ we need to impose the relation \eqref{STC}. 
By approximating the left and right values of the current with, respectively, backward and forward second order finite differences, i.e.
$$
\begin{aligned}
&j^1\approx  \frac{n^{N-2}-4n^{N-1}+3n^{N}}{2\Delta x} - \beta\cE n^{N} ,
\\[4pt]
&j^2\approx  \frac{-3n^{N+1}+4n^{N+2}-n^{N+3}}{2\Delta x} - \beta\cE n^{N+1} ,
\end{aligned}
$$
we first write the flux conservation $j^1 = j^2$ as follows:
\BE
\frac{n^{N-2}-4n^{N-1}+3n^{N}}{2\Delta x} - \beta \cE n^{N} 
-  \frac{-3n^{N+1}+4n^{N+2}-n^{N+3}}{2\Delta x} + \beta\cE n^{N+1}  =0.
\EE
Then, the transmission condition \eqref{STC} can be written (by using, e.g., $j^2$ for $j$)
\BE
n^{N}-n^{N+1} - \vartheta \left( \frac{-3n^{N+1}+4n^{N+2}-n^{N+3}}{2\Delta x} - \beta\cE n^{N+1} \right) = 0.
\EE
The interpolation constant $\vartheta$ is computed numerically from \eqref{qMB}, as described above, 
by means of standard integration routines.
\par
Using the above described model  we have computed the conductance as a function of the top gate voltage $V_\mathit{tg}$ 
for different values of the back gate voltage $V_\mathit{bg}$ and for different values of temperature.
The values of the physical parameters used in our simulations are similar to those of the device described in 
Ref.\ \cite{YK2009}, namely $L = 4\,\mu\mathrm{m}$, $D = 0.05\,\mu\mathrm{m}$, $\tau = 0.075\,\mathrm{ps}$.
The bias voltage applied at the contacts is $V_\mathit{bias} = 0.001\,\mathrm{V}$ and the contact width is $1\,\mu\mathrm{m}$
The constant $c_{\mathit{tg}}$ (see Eq.\ \eqref{Ehdef}) has been used as a tuning parameter and has been set to $0.05$. 
Then, the value of  $c_{\mathit{bg}}$ has been set to $12.8\, c_{\mathit{tg}}$, so to maintain the same ratio between the 
corresponding capacitive constants as in Ref.\ \cite{YK2009}.
The values of the densities at the contacts, $n_l$ and $n_r$, has been simply set equal to the background density, 
that is
$$
   n_l =  n_r = n_0\,\phi_2(\beta q c_{\mathit{bg}}V_\mathit{bg})
$$
(however, more refined models for the boundary densities could also be considered, see e.g.\ Ref.\ \cite{LJG2014}).
Finally, the conductance, that is the ratio between the total electric current flowing through the device and the bias voltage, 
is expressed in our graphs in units of the ``quantum of conductance'' $q^2/h$.
The total current, is computed from the current density $j$ by assuming a device effective width of $1\,\mu\mathrm{m}$.
\begin{figure}
\begin{center}
\includegraphics[width=\linewidth]{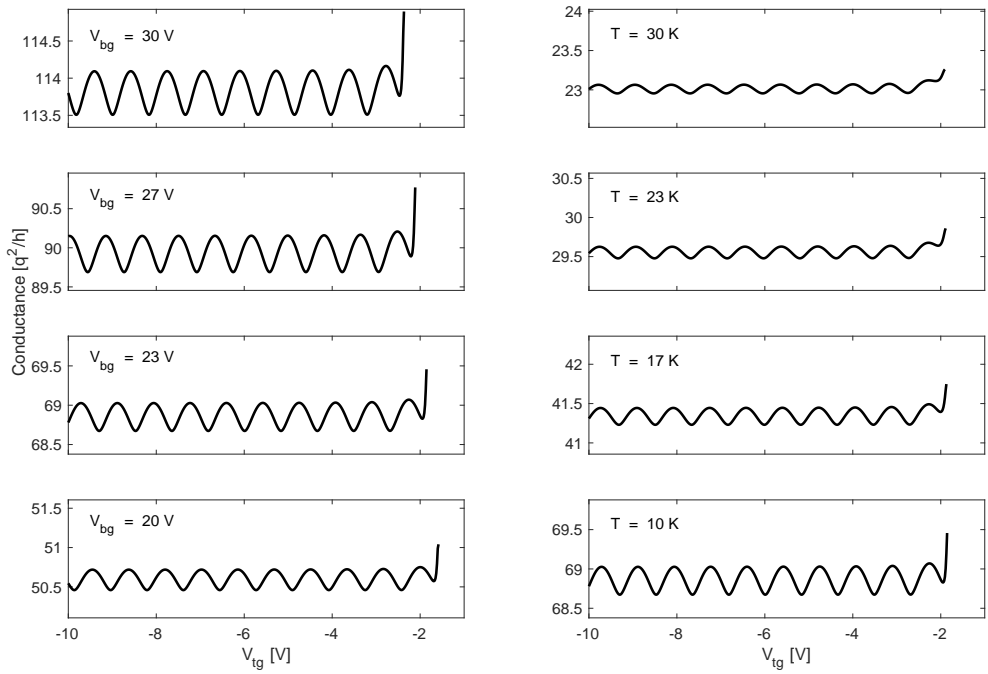}  
\caption{Conductance as a function of the top gate voltage $V_\mathit{tg}$ for different
values of the back gate (left column) and for different values of the temperature (right column).
In the left plots, the temperature is fixed at $T = 10\,\mathrm{K}$ while, in the right plots, the back gate voltage
is fixed at $V_\mathit{tg} = 23\, \mathrm{V}$.
}
\label{figura3}
\end{center}
\end{figure}
%

\par
The results of the numerical simulations are reported in Figure \ref{figura3}. 
In the left column we show plots of the conductance as a function of the top gate voltage $V_\mathit{tg}$ for different
values of the back gate, at a fixed temperature $T = 10\,\mathrm{K}$.
We see that, as long as 
$E_h = -qc_{\mathit{bg}} V_{\mathit{bg}}  -qc_{\mathit{tg}} V_{\mathit{tg}} > 0$ (correponding to the ``n-p-n'' case),
the conductance shows Fabry-Perot-like oscillations whose amplitude and period of the same order of those 
reported in the experimental literature \cite{YK2009}.
The oscillations are followed by sudden rise as $E_h$  approaches 0 (which is also observed in the experiments \cite{Huard2007,YK2009}).
The shift towards the left of the point  $E_h = 0$ for increasing  $V_{\mathit{bg}}$
is also typical of experimental observation and is an obvious consequence of the relation between 
$E_h$, $V_{\mathit{bg}}$ and $V_{\mathit{tg}}$. 
Moreover, still in accordance with experimental measurements, the conductance oscillates around around a mean value that increases, approximately linearly, with $V_\mathit{bg}$.
\par
In the right column are shown plots of the conductance as a function of the top gate voltage $V_\mathit{tg}$ for different
values of the temperature, at a fixed back gate voltage $V_{\mathit{bg}}  = 23\,\mathrm{V}$.
We see that the amplitude of the Fabry-Perot oscillations strongly decreases by increasing temperature, 
as reported from experiments \cite{YK2009}.
A simple explanation of the latter phenomenon is apparent from Fig.\ \ref{figura2}: decreasing the temperature makes the Maxwellian
narrower thus making the interpolation constant $\vartheta$ more sensitive to the variations of the transmission coefficient.
\section{Conclusions}
\label{S5}
In this paper we have shown how the theory of diffusive transmission boundary conditions at a quantum interface, 
developed in Ref.\ \cite{BN2018} and summarised in Section \ref{S2} of the present paper, 
can be applied for the numerical simulation of a heterojunction graphene device. 
To this aim, we also had to expand the theoretical part in order to simplify the kinetic step, represented by the solution of a four-fold Milne problem. 
In fact, the asymptotic densities associated to the solution of such Milne problem provide the interpolation constant, which is the key ingredient in the 
formulation of the transmission conditions.
This has been done in Sec.\ \ref{S3}, where explicit expressions of the asymptotic densities have been obtained assuming of very short collision times.
\par
The material developed in Secs.\ \ref{S2} and \ref{S3} is then used in Sec.\ \ref{S4} to set up a mathematical model of a generic ``n-p-n'' graphene device, 
with some additional simplifying assumptions (above all the fact that the devices works in a regime where only the electron population is relevant and where 
the Maxwell-Boltzmann statistics can be used instead of the Fermi-Dirac one).
Indeed, our simulations are intended to illustrate the method and its potentialities, rather than to faithfully reproduce a specific device in a specific regime.
\par
In spite of all these simplifications, the numerical simulations reported in Sec.\ \ref{S4.1} show that the model is able to reproduce some important features
of the electron transport in n-p-n graphene heterojunctions (see Figure \ref{figura3}).
In particular, we observe the Fabry-Perot oscillations of the conductance, which are the signature of quantum interference inside the barrier and of the chiral
nature of the electrons in graphene. 
Such oscillations have the expected behaviour with respect to the variations of the gate voltages as well as to the variations of temperature.
Still in accordance with laboratory observations, the conductance has a sudden increase when the barrier height 
$E_h$ approaches zero.
\par
A feature that our simulations are unable to describe is the behaviour of the conductance when the barrier height 
$E_h$ enters the negative range (the ``n-n-n'' case). 
In fact, when $E_h$ becomes negative, experiments show that the conductance, after the sudden jump described above, 
keeps on increasing 
(more slowly) and oscillations disappear. 
Instead, our simulations predict a new decrease of the conductance 
and new oscillations (not shown in the figure). 
We believe that this fault is due to the ideal model of perfectly sharp barrier that we adopted. 
On the other hand, the theoretical models adopted in Refs.\ \cite{Huard2007,YK2009} are also unable to describe such behaviour
(neither are the other models in literature we are aware of). 
Probably, more refined descriptions of the barrier, requiring however a numerical solution of the scattering problem \eqref{SE}, could permit
to reproduce correctly the behaviour of the device across the entire range of $E_h$.
\par
Discrepancies of the overall values of the conductivity with respect to those observed in experiments 
are mainly due to the oversimplified model for the mobility that we are using, which is not suited to accurately describe the great 
variety of experimental devices.
However, as remarked above, since the aim of this paper is mainly to illustrate the mathematical method of quantum transmission conditions, rather 
than to simulate a particular device, we preferred to use a simple model for the bulk transport and to focus on the treatment 
of the quantum interface.
Of course, a more detailed description of the electron scattering \cite{CocoEtAl19,Morandi11,MajoranaNastasiRomano19},
more refined expressions for the mobility \cite{NastasiRomano19,NR20b}, a self-consistent potential model \cite{NR20a,NR20b} 
or even quantum drift-diffusion equations \cite{LucaRomano19,Romano07,ZB11}, 
could be used to improve the model (see also Ref.\ \cite{CMR2020} for a general reference).
\section*{Acknowledgements}
All authors acknowledge support from French-Italian University "Galileo" project 
{\it Classical and quantum kinetic models and their hydrodynamic limits: theoretical and applied aspects} (Ref. G18\_296).
G.N. acknowledges support from ``Progetto Giovani'' of the italian National Group for Mathematical Physics - GNFM 2019 {\it Modelli matematici, numerici e simulazione del trasporto di cariche e fononi nel grafene}.
Support is also acknowledged from Universit\`a degli Studi di Catania, "Piano della Ricerca 2016/2018 Linea di intervento 2".

\end{document}